\documentclass[twocolumn,showpacs,pre]{revtex4}
\usepackage{amsmath}
\usepackage{graphicx}%
\usepackage{color}
\usepackage{amsfonts}

\begin{document}

\title{Diverging length scale of the inhomogeneous mode-coupling theory:
a numerical investigation}

\author{Grzegorz Szamel and Elijah Flenner}
\affiliation{Department of Chemistry, Colorado State University, Fort Collins, CO 80523}
\date{\today}

\begin{abstract}
Biroli \textit{et al.}'s extension of the standard mode-coupling theory to inhomogeneous equilibrium states
[Phys. Rev. Lett. \textbf{97}, 195701 (2006)] allowed them to identify a characteristic length scale that
diverges upon approaching the mode-coupling transition. 
We present a numerical investigation of this length scale. To this end we derive and numerically
solve equations of motion for coefficients in the small $q$ expansion of the dynamic susceptibility 
$\chi_{\mathbf{q}}(\mathbf{k};t)$ that describes the change of the system's dynamics due to an external
inhomogeneous potential. We study the dependence of the characteristic length scale on time, wave-vector,
and on the distance from the mode-coupling transition.  We verify scaling predictions of Biroli \textit{et al.} 
In addition, we find that the numerical value of the diverging length scale qualitatively agrees with
lengths obtained from four-point correlation functions. 
We show that the diverging length scale has very weak $k$ dependence, which contrasts with very 
strong $k$ dependence of the $q\to 0$ limit of the susceptibility, $\chi_{\mathbf{q}=0}(\mathbf{k};t)$.
Finally, we compare the diverging length obtained from the small $q$ expansion to that resulting from an 
isotropic approximation applied to the equation of motion for the dynamic susceptibility 
$\chi_{\mathbf{q}}(\mathbf{k};t)$.
\end{abstract}


\maketitle


\section{Introduction}
As a liquid is cooled, its dynamics not only gets slower but also becomes increasingly heterogeneous 
\cite{Ediger2000,Richert2002,Andersen2005}.
Moreover, the characteristic size of regions with dynamics both significantly faster and 
significantly slower than the average dynamics grows upon cooling. 
This observation has led to the definition of a dynamic correlation length that measures the 
size of these so-called dynamic heterogeneneities. 
The dynamic correlation length was defined in terms of a four-point correlation function \cite{Lacevic2002}
or a corresponding four-point structure factor 
\cite{Lacevic2003,Berthier2004,Toninelli2005,Berthier2007p2}.

While various four-point functions can readily be obtained from simulations (albeit they typically require
more computational effort than the familiar two-point functions), they are difficult to access experimentally.
To the best of our knowledge, four-point functions have been obtained directly only from
experiments on granular systems \cite{Lechenault2008a,Lechenault2008b}. In a remarkable development,
Berthier \textit{et al.} \cite{Berthier2005} showed that derivatives of standard two-point 
functions with respect to thermodynamic variables like, \textit{e.g.}, density or temperature, could be
related to integrals of three-point correlation functions. This opened a door to experimental
investigations of the overall degree (the strength) of dynamic heterogeneity upon approaching the glass transition
\cite{Dalle-Ferrier2007}. It should be emphasized, however, that Berthier \textit{et al.} could obtain
the dynamic correlation length characterizing the spatial extent of dynamic heterogeneneities only by using 
additional assumptions that related the integrals of various three-point functions to 
characteristic length scales exhibited by these functions.

While theoretical understanding of the derivatives of two-point functions with respect to 
thermodynamic variables is limited, Biroli \textit{et al.} \cite{Biroli2006} showed that 
the mode-coupling theory could be used to analyze a closely related quantity, a three-point susceptibility 
$\chi_{\textbf{q}}(\textbf{k};t)$, which describes the change of the intermediate
scattering function due to an inhomogeneous external potential. The advantage of this approach is
that it allows one to evaluate a characteristic length scale which, up to that time, had remained
hidden within the well-known mode-coupling framework. This was possible due to the fact that 
Biroli \textit{et al.} considered a \textit{non-uniform} 
external perturbation rather than a uniform change of density or temperature. 

To analyze the three-point susceptibility Biroli \textit{et al.} extended the standard mode-coupling
theory to describe the time evolution of the intermediate scattering function of a system under the influence
of an inhomogeneous external potential. They defined the three-point susceptibility 
$\chi_{\textbf{q}}(\textbf{k};t)$ as a derivative of the intermediate scattering function 
with respect to a Fourier component of the external potential, $U(\mathbf{q})$. 
Biroli \textit{et al.} showed that upon approaching the mode-coupling transition both the $q\to 0$ limit
of the three-point susceptibility and a characteristic length defined through the small $q$-dependence of 
$\chi_{\textbf{q}}(\textbf{k};t)$ diverge. Moreover, they derived scaling predictions
for the time-dependence of the characteristic length, and they found that this length grows as
$t^{a/2}$ in the early $\beta$ regime (here $a$ is the mode-coupling exponent describing approach 
of the intermediate scattering function to its plateau value) and then saturates in the 
late $\beta$ and $\alpha$ regimes. This was contrasted with the time dependence of the $q\to 0$ limit
of the three-point susceptibility which grows as $t^a$ and $t^b$ in the 
early and late $\beta$ regimes, respectively (here $b$ is the so-called von Schweidler exponent 
of the mode-coupling theory describing departure of the intermediate scattering function from its plateau value), 
peaks around the $\alpha$ relaxation time, and then decays to zero. 
The strikingly different time-dependence of the characteristic length and
the $q\to 0$ limit of the three-point susceptibility was interpreted as an indication of 
changing fractal dimension of dynamic heterogeneities. 

Biroli \textit{et al.} derived only scaling predictions for the three-point susceptibility 
$\chi_{\textbf{q}}(\textbf{k};t)$. They verified some of these predictions through a numerical analysis
of a schematic model that completely disregards $k$ dependence. Here we present a numerical investigation of the 
small $q$ behavior of the three-point susceptibility and the associated characteristic length.
We focus on the time and $k$ dependence of the length, and on its dependence on the 
distance to the mode-coupling transition. 

We start with a definition of the three-point susceptibility  $\chi_{\textbf{q}}(\textbf{k};t)$ 
in Sec.~\ref{sec:def}.
Next, we postulate an expansion of the three-point susceptibility  $\chi_{\textbf{q}}(\textbf{k};t)$
in powers of $\mathbf{q}$ and derive equations of motion for the first few coefficients in this expansion
(see Sec.~\ref{sec:expand}). We also present an alternative approach to a numerical evaluation
of the characteristic length which is based on an isotropic approximation 
to the equation of motion (see Sec.~\ref{sec:isotropic}).
Next, in Secs.~\ref{sec:numerical} and \ref{sec:numericaliso} we describe the results of the numerical
calculations based on the small $q$ expansion and the isotropic approximation, respectively. 
We finish  with a summary and conclusions in Sec.~\ref{sec:conclusions}.

\section{Three-point susceptibility $\chi_{\textbf{q}}(\textbf{k};t)$}\label{sec:def}

To obtain the equation of motion for the three-point susceptibility Biroli \textit{et al.} \cite{Biroli2006}
considered a Newtonian fluid subject to a periodic in space external potential, derived
mode-coupling equation of motion for the intermediate scattering function of this inhomogeneous system, 
and then differentiated this 
equation with respect to the external potential. Subsequently, one of us has derived the equation of motion for 
the same three-point susceptibility for a Brownian system \cite{SzamelBDMCT}. 
Not surprisingly, the overdamped limit of the equation 
of motion derived by Biroli \textit{et al.} coincides with the equation of motion derived starting directly from
Brownian dynamics. In this work we will use the latter equation.

For a system subject to a non-uniform external potential the intermediate scattering function is not
diagonal in the wave-vector,
\begin{equation}\label{eq:Fktdef}
F(\mathbf{k}_1,\mathbf{k}_2;t) = \frac{1}{N} \left< \rho(\mathbf{k}_1;t) \rho(-\mathbf{k}_2) \right>_U.
\end{equation}
Here $\rho(\mathbf{k}_1;t)$ is the Fourier transform of the microscopic density,
\begin{equation}\label{rhodef}
\rho(\mathbf{k}_1;t) = \sum_j e^{-i \mathbf{k}_1\cdot\mathbf{r}_j(t)}
\end{equation}
with $\mathbf{r}_j(t)$ being the position of the $j$th particle at time $t$. Furthermore, in Eq. (\ref{eq:Fktdef})
$\rho(\mathbf{k}_1)\equiv \rho(\mathbf{k}_1;t=0)$ and $\left< ... \right>_U$ denotes the equilibrium average
for a system subject to a static non-uniform external potential $U$.

The three-point susceptibility $\chi_{\textbf{q}}(\mathbf{k};t)$ is defined through an expansion of the 
intermediate scattering function in powers of a harmonic external potential 
$U_{\mathbf{q}} = U_0 e^{-i\mathbf{q}\cdot\mathbf{r}}$,
\begin{equation}\label{eq:chidef}
F(\mathbf{k},\mathbf{k}_1;t) = F(k;t)\delta_{\mathbf{k},\mathbf{k}_1} + 
\chi_{\textbf{q}}(\mathbf{k};t) \left(-\beta U_0 \right) \delta_{\mathbf{k}+\mathbf{q},\mathbf{k}_1} + ... .
\end{equation}

For a system with Brownian dynamics, the equation of motion for the
three-point susceptibility $\chi_{\mathbf{q}}(\mathbf{k};t)$ has the following form:
\begin{eqnarray}\label{eq:motion}
&& \partial_t \chi_{\mathbf{q}}(\mathbf{k};t) +\frac{D_0 k^2}{S(k)} \chi_{\mathbf{q}}(\mathbf{k};t) 
\nonumber \\ &&
+ \int_0^t dt^\prime M^{\mathrm{irr}}(k;t-t^\prime) \partial_{t^\prime} \chi_{\mathbf{q}}(\mathbf{k};t^\prime)
\nonumber \\ &&
+ \int_0^t dt^\prime 
M^{\chi}_{\mathbf{q}}(\mathbf{k};t-t^\prime)\partial_{t^\prime} F(|\mathbf{k} + \mathbf{q}|;t^\prime)
\nonumber\\ &=& 
\mathcal{S}_{\mathbf{q}}(\mathbf{k};t) 
\end{eqnarray}
In Eq.~\eqref{eq:motion} 
$D_0$ is the diffusion coefficient of an isolated particle, $S(k)$ denotes the static structure factor, 
$M^{\mathrm{irr}}(k;t)$ is the irreducible memory function of mode-coupling theory,
\begin{eqnarray}\label{Mirr}
&& M^{\mathrm{irr}}(k;t) = \\ \nonumber &&  
\frac{n D_0}{2} \int \frac{d \mathbf{k}_1}{(2 \pi)^2} [v_{\mathbf{k}}(\mathbf{k}_1,\mathbf{k}-\mathbf{k}_1)]^2 
F(k_1;t)F(|\mathbf{k}-\mathbf{k}_1|;t) ,
\end{eqnarray}
and $M^{\chi}(\mathbf{q},k;t)$ is defined as follows,
\begin{eqnarray}\label{Mchi}
&& M^{\chi}_{\mathbf{q}}(\mathbf{k};t) = \frac{n D_0 k}{|\mathbf{k} + \mathbf{q}|} 
\int \frac{d \mathbf{k}_1}{(2 \pi)^3} 
v_{\mathbf{k}}(\mathbf{k}_1,\mathbf{k}-\mathbf{k}_1) 
\\ && \nonumber \times 
\chi_{\mathbf{q}}(\mathbf{k}_1;t)  F(|\mathbf{k}-\mathbf{k}_1|;t)
v_{\mathbf{k} + \mathbf{q}}(\mathbf{k}_1 + \mathbf{q},\mathbf{k}-\mathbf{k}_1) .
\end{eqnarray}
In Eqs. (\ref{Mirr}-\ref{Mchi}) 
$v_{\mathbf{k}}(\mathbf{k}_1,\mathbf{k}_2) = 
\hat{\mathbf{k}} \cdot (\mathbf{k}_1 c(k_1) + \mathbf{k}_2 c(k_2))$ with $\hat{\mathbf{k}} = \mathbf{k}/k$, 
and $n$ is the density. 
The source term in Eq. (\ref{eq:motion}), $\mathcal{S}_{\mathbf{q}}(\mathbf{k};t)$, is given by
\begin{eqnarray} \label{eq:source}
\lefteqn{\mathcal{S}_{\mathbf{q}}(\mathbf{k};t)  = } \\ \nonumber 
&& D_0 k^2 S(q) \left( 1 - \frac{\mathbf{k} \cdot 
(\mathbf{k} + \mathbf{q})}{k^2 S(|\mathbf{k} + \mathbf{q}|)} \right) F(|\mathbf{k} + \mathbf{q}|;t) 
\nonumber \\ && \nonumber
+S(q) \int_0^t dt^\prime M^{\mathrm{irr}}(k;t-t^\prime) \frac{\mathbf{k} \cdot 
(\mathbf{k}+\mathbf{q})}{|\mathbf{k} + \mathbf{q}|^2} \partial_{t^\prime} F(|\mathbf{k} + \mathbf{q}|;t^\prime).
\end{eqnarray}
Finally, the  initial condition for the three-point susceptibility 
is $\chi_{\mathbf{q}}(\mathbf{k};t=0) = S(k)S(q)S(|\mathbf{k}+\mathbf{q}|)$. This 
form of the initial condition is obtained by applying a convolution approximation to the exact expression
for the initial condition, which involves a three-particle correlation function. One should note that 
the same convolution approximation is used in the derivation of the mode-coupling equation of motion 
(both in a uniform and a non-uniform equilibrium state).

It should be emphasized at this point that solving Eq. (\ref{eq:motion}) numerically is considerably
more involved than solving the uniform equilibrium mode-coupling equations 
\cite{Fuchs1991,Miyazaki,Flenner2005sim},
and, to the best of our knowledge, has never been attempted. 
The reason is that while $\mathbf{q}$ is a parameter in Eq. (\ref{eq:motion}), non-zero value of
$\mathbf{q}$ breaks rotational symmetry. Thus, $\chi_{\mathbf{q}}(\mathbf{k};t)$ depends not only
on $k=|\mathbf{k}|$ and $q=|\mathbf{q}|$, but also on the angle between $\mathbf{k}$ and $\mathbf{q}$.
Most importantly, the latter angle is an independent variable, rather than a parameter in Eq. \eqref{eq:motion}.


\section{Expansion of $\chi_{\mathbf{q}}(\mathbf{k};t)$}\label{sec:expand}

\subsection{Preliminaries}

In this work we focus on the characteristic length defined through the small $q$-dependence of 
the three-point susceptibility $\chi_{\mathbf{q}}(\mathbf{k};t)$.  
Thus, to calculate this length we only need to 
obtain the small $q$ behavior of the susceptibility. We postulate the following 
expansion of $\chi_{\mathbf{q}}(\mathbf{k};t)$ in powers of $\mathbf{q}$,
\begin{eqnarray} \label{eq:expansion}
\chi_{\mathbf{q}}(\mathbf{k};t) & = & \chi^{(0)}(k;t) + 
\sum_\alpha q_\alpha \left[\frac{\partial 
\chi_{\mathbf{q}}(\mathbf{k};t)}{\partial q_\alpha}\right]_{\mathbf{q}=0}
\nonumber \\ && 
+\sum_{\alpha \beta} \frac{q_\alpha q_\beta}{2} 
\left[\frac{\partial^2 \chi_{\mathbf{q}}(\mathbf{k};t)}{\partial q_\alpha \partial q_\beta}\right]_{\mathbf{q}=0}
+ \ldots
\nonumber \\ 
&=& \chi^{(0)}(k;t) + \hat{\mathbf{k}} \cdot \mathbf{q} \chi^{(1)}(k;t) 
\nonumber \\ &&
+ \mathbf{q}^2 \chi^{(2)}(k;t) 
\nonumber \\ &&
+ \left(3 \left(\hat{\mathbf{k}}\cdot\mathbf{q} \right)^2- \mathbf{q}^2 \right) \chi_{\mathrm{tl}}^{(2)}(k;t) 
+ \ldots 
\end{eqnarray}
where the second equality follows from symmetry considerations and quantities $\chi^{(1)}$, $\chi^{(2)}$,
and $\chi_{\mathrm{tl}}^{(2)}$ are defined as follows:
\begin{eqnarray}\label{eq:chi0def}
\chi^{(1)}(k;t) & = & \sum_\alpha \hat{k}_\alpha \left. 
\frac{\partial \chi_{\mathbf{q}}(\mathbf{k};t)}{\partial q_\alpha} \right|_{\mathbf{q} = 0}
\\ \label{eq:chi1def}
\chi^{(2)}(k;t) & = &  \frac{1}{6} \sum_{\alpha} 
\left. \frac{\partial^2 \chi_{\mathbf{q}}(\mathbf{k};t)}{\partial q_\alpha \partial q_\alpha} \right|_{\mathbf{q}
= 0}
\\ \label{eq:chi2def}
\chi_{\mathrm{tl}}^{(2)}(k;t) & = & \frac{1}{4} \sum_{\alpha \beta} 
\left( \hat{k}_\alpha \hat{k}_\beta  - \frac{1}{3} \delta_{\alpha \beta} \right)
\left. \frac{\partial^2 \chi_{\mathbf{q}}(\mathbf{k};t)}{\partial q_\alpha q_\beta}\right|_{\mathbf{q} = 0}
\end{eqnarray}

We show in Sec.~\ref{sec:numerical} that the first order term, $\chi^{(1)}(k;t)$, does not
lead to a diverging characteristic length scale. Furthermore, we show that the second
order term originating from  the trace of the second derivative of $\chi_{\mathbf{q}}(\mathbf{k};t)$, 
$\chi^{(2)}(k;t)$, leads to a diverging characteristic length scale. Finally, it can be shown that the
second order term originating from the symmetric traceless part of the second derivative of 
$\chi_{\mathbf{q}}(\mathbf{k};t)$, $\chi_{\mathrm{tl}}^{(2)}(k;t)$, does not lead to a diverging characteristic 
length scale thus we omit the equation of motion for $\chi_{\mathrm{tl}}^{(2)}(k;t)$ for sake of space a clarity. 

\subsection{Zeroth order coefficient $\chi^{(0)}(k;t)$}

To get the equation of motion for $\chi^{(0)}(k;t)$ we need to take $\mathbf{q}\to 0$ limit in all terms in
Eq. (\ref{eq:motion}). In this way we obtain the following equation,
\begin{eqnarray}\label{eq:chizero}
&&\partial_t \chi^{(0)}(k;t) + D_0 \frac{k^2}{S(k)} \chi^{(0)}(k;t)
\nonumber \\ &&
+ \int_0^t dt^\prime M^{\mathrm{irr}}(k;t-t^\prime) \partial_{t^\prime} \chi^{(0)}(k;t^\prime)
\nonumber \\ && 
+ \int_0^t dt^\prime M^{\chi}_0(k;t-t^\prime)
\partial_{t^\prime}F(k;t^\prime)
\nonumber\\
&=& 
n D_0 k^2 S(0) c(k) F(k;t) 
\nonumber \\ &&
+ S(0) \int_0^t dt^\prime M^{\mathrm{irr}}(k;t-t^\prime) \partial_{t^\prime} F(k;t^\prime)
\end{eqnarray}
where
\begin{eqnarray}\label{eq:Mchi0}
M^{\chi}_0(k;t) & = &
 n D_0  \int \frac{d \mathbf{k}_1}{(2 \pi)^3} [v_{\mathbf{k}}(\mathbf{k}_1,\mathbf{k}-\mathbf{k}_1)]^2 
\nonumber \\ && \times 
\chi^{(0)}(k_1;t) F(|\mathbf{k}-\mathbf{k}_1|;t).
\end{eqnarray}
Furthermore, taking $\mathbf{q}\to 0$ limit of the initial condition 
$\chi_{\mathbf{q}}(\mathbf{k};t=0)=S(k)S(q)S(|\mathbf{k}+\mathbf{q}|)$ we obtain the initial condition
for $\chi^{(0)}(k;t)$, $\chi^{(0)}(k;t=0)=S(0)S(k)^2$.

The zeroth order coefficient, $\chi^{(0)}(k;t)$, satisfies essentially
the same equation of motion as three point susceptibility $\chi_n(k;t)$, which is defined as the 
density derivative of the intermediate scattering function \cite{Szamel2009}. 
This was expected: in the long wavelength,
$\mathbf{q}\to 0$, limit the derivative with respect to the external potential differs from the 
derivative with respect to the density by a thermodynamic factor proportional to
$\left(\partial n /\partial \beta \mu\right)_T$. 

The equation of motion \eqref{eq:chizero} can be solved using 
the static structure factor $S(k)$ and the dynamic scattering
function $F(k;t)$ as input. We calculate $F(k;t)$ using the mode-coupling theory; the equation of motion
for $F(k;t)$ reads,
\begin{eqnarray}\label{eq:mct}
\partial_t F(k;t) &+& \frac{D_0 k^2}{S(k)} F(k;t) \nonumber \\ &+&  
\int_0^t dt^\prime M^{\mathrm{irr}}(k;t-t^\prime) \partial_{t^\prime} F(k;t^\prime) = 0,
\end{eqnarray}
where $M^{\mathrm{irr}}$ is the irreducible memory function given by Eq. (\ref{Mirr}). 

\subsection{First order coefficient $\chi^{(1)}(k;t)$}

To get the equation of motion for $\chi^{(1)}(k;t)$ we need to expand all the terms in Eq. (\ref{eq:motion})
in powers of $\mathbf{q}$ and then to collect terms 
linear in $\mathbf{q}$. After exploiting rotational symmetry we get the following equation of motion
\begin{eqnarray} \label{eq:chione}
&&\partial_t \chi^{(1)}(k;t) + \frac{D_0 k^2}{S(k)}\chi^{(1)}(k;t) 
\\ && +
\int_0^t dt^\prime M^{\mathrm{irr}}(k;t-t^\prime) \partial_{t^\prime} \chi^{(1)}(k;t^\prime) 
\nonumber \\ && 
+ \int_0^t dt^\prime M_1^{\chi}(k;t-t^\prime) \partial_{t^\prime} F(k;t^\prime)
\nonumber \\ && 
+ \int_0^t dt^\prime M_{0}^{\chi}(k;t-t^\prime)\partial_{t^\prime} \partial_k F(k;t^\prime) 
\nonumber \\ &=&
\frac{D_0 S(0) F(k;t) k^2}{S(k)}\left[ \frac{1}{S(k)} \frac{d S(k)}{dk} - \frac{1}{k} \right] 
\nonumber \\ &&
+  n D_0 k^2 S(0) c(k) \partial_k F(k;t) 
\nonumber \\ &&
+S(0) \int_0^t dt^\prime M^{\mathrm{irr}}(k;t-t^\prime) \partial_{t^\prime} \partial_k F(k;t^\prime) 
\nonumber \\ &&
- \frac{S(0)}{k} \int_0^t dt^\prime M^{\mathrm{irr}}(k;t-t^\prime) \partial_{t^\prime} F(k;t^\prime) \nonumber,
\end{eqnarray}
where
\begin{eqnarray}
\lefteqn{ M_1^{\chi}(k;t)  = } \nonumber \\ && 
n D_0 
\int \frac{d \mathbf{k}_1}{(2 \pi)^3} [v_{\mathbf{k}}(\mathbf{k}_1,\mathbf{k}-\mathbf{k}_1)]^2 
\frac{\mathbf{k}\cdot\mathbf{k}_1}{k k_1} 
\nonumber \\ && \times 
F(|\mathbf{k}-\mathbf{k}_1|;t) \chi^{(1)}(k_1;t)
\nonumber \\ && 
+n D_0 \int \frac{d \mathbf{k}_1}{(2 \pi)^3} v_{\mathbf{k}}(\mathbf{k}_1,\mathbf{k}-\mathbf{k}_1) 
F(|\mathbf{k}-\mathbf{k}_1|;t) \chi^{(0)}(k_1;t)
\nonumber \\ && \times 
\left\{   c(k_1) + \frac{ [\mathbf{k} \cdot \mathbf{k}_1]^2}{k^2 k_1} \frac{d c(k_1)}{d k_1} 
- \frac{v_{\mathbf{k}}(\mathbf{k}_1,\mathbf{k}-\mathbf{k}_1)}{k} \right\}.
\end{eqnarray}
Furthermore, we obtain the following expression for the initial condition for $\chi^{(1)}(k;t)$,
$\chi^{(1)}(k;t=0) = S(k) S(0) dS(k)/dk$.

To solve Eq.~\eqref{eq:chione} we need the equation of motion for the partial derivative of the 
intermediate scattering function with respect to the wave-vector, $\partial_k F(k;t)$. This 
equation can be derived from the mode-coupling equation \eqref{eq:mct}:
\begin{eqnarray}\label{eq:dkF}
&& \partial_t \partial_k F(k;t) + \frac{D_0 k^2}{S(k)} \partial_k F(k;t)
\nonumber \\ && 
+ \frac{D_0 k}{S(k)}\left[2- \frac{k}{S(k)} \partial_k S(k) \right] F(k;t) 
\nonumber \\ &&
 + \int_0^t dt^\prime M^{\mathrm{irr}}(k;t-t^\prime) \partial_{t^\prime} \partial_k F(k;t^\prime) \nonumber \\
&&+  \int_0^t M_1^{k1}(k;t-t^\prime) \partial_{t^\prime} F(k;t^\prime) \nonumber\\
&&+  \int_0^t d t^\prime M_2^{k1}(k;t-t^\prime) \partial_{t^\prime} 
F(k;t^\prime) = 0
\end{eqnarray}
where
\begin{eqnarray}\label{eq:Mk1}
\lefteqn{M_1^{k1}  = } \\ \nonumber && 
n D_0 \int \frac{d \mathbf{k}_1}{(2 \pi)^3}\left[ v_{\mathbf{k}}(\mathbf{k}_1,\mathbf{p}) \right]
\left\{ c(p) + \frac{[\mathbf{k} \cdot \mathbf{p}]^2}{k^2 p}  
\frac{d c(p)}{d p} \right\} \nonumber \\ \nonumber
&&\times F(k_1;t) F(p;t),
\end{eqnarray}
and 
\begin{eqnarray}\label{eq:Mk2}
\lefteqn{M_2^{k1}(k;t)  = } \\ \nonumber && 
\frac{n D_0}{2} \int \frac{d \mathbf{k}_1}{(2 \pi)^3}[v_{\mathbf{k}}(\mathbf{k}_1,\mathbf{p})]^2
 F(k_1;t) \frac{\mathbf{k} \cdot \mathbf{p}}{k p} \partial_{p} F(p;t).
\end{eqnarray}
In Eqs. (\ref{eq:Mk1}-\ref{eq:Mk2}) $\mathbf{p}=\mathbf{k}-\mathbf{k}_1$ and $p=|\mathbf{k}-\mathbf{k}_1|$.

\subsection{Second order coefficient $\chi^{(2)}(k;t)$}

For symmetry reasons, there are two linearly independent second order coefficients, the coefficient proportional
to the trace of the matrix of second derivatives of $\chi_{\mathbf{q}}(\mathbf{k};t)$, $\chi^{(2)}(k;t)$, and
the coefficient proportional to the symmetric traceless part of the matrix of second derivatives of 
$\chi_{\mathbf{q}}(\mathbf{k};t)$, $\chi^{(2)}_{tl}(k;t)$. 
It can be shown that only the former coefficient leads to a characteristic
length that diverges upon approaching the mode-coupling transition. Therefore, since the focus of this work
is the diverging characteristic length,
we will only give the equation of motion for $\chi^{(2)}(k;t)$. 
By expanding equation of motion (\ref{eq:motion}) in powers of $\mathbf{q}$, collecting the second order
terms and taking a trace of the corresponding tensorial equation of motion we can derive the following
equation of motion for  $\chi^{(2)}(k;t)$:
\begin{eqnarray}\label{eq:chitwo}
&&\partial_t \chi^{(2)}(k;t)
+ D_0 \frac{k^2}{S(k)} \chi^{(2)}(k;t)
\nonumber \\ &&
+\int_0^t dt^\prime M^{\mathrm{irr}}(k;t-t^\prime) \partial_{t^\prime} \chi^{(2)}(k;t^\prime)
 \nonumber \\ && 
 +\int_0^t dt^\prime 
M_2^\chi(k;t-t^\prime)  \partial_{t^\prime} F(k;t^\prime)
 \nonumber \\ &=& 
- \frac{1}{6} \int_0^t dt^\prime 
M^\chi_0(k;t-t^\prime)
\partial_{t^\prime} \partial^2_k F(k;t^\prime)   
 \nonumber \\ && 
 -\frac{1}{3k} \int_0^t dt^\prime 
M^\chi_0(k;t-t^\prime) \partial_{t^\prime} \partial_k F(k;t^\prime) 
 \nonumber \\ &&
-\frac{2}{3} \int_0^t dt^\prime 
M^\chi_1(k;t-t^\prime) \partial_{t^\prime} \partial_k F(k;t^\prime)
  \nonumber \\ && 
 -\frac{2}{3} \int_0^t dt^\prime 
 M^\chi_3(k;t-t^\prime) \partial_{t^\prime} F(k;t^\prime)
 \nonumber \\ &&
 +\mathcal{S}^{(2)}(k;t),
 \end{eqnarray}
 where
 \begin{eqnarray}
 \lefteqn{\mathcal{S}^{(2)}(k;t) = } \\ &&
 \left[ \left. \frac{d^2 S(q)}{dq^2} \right|_{q=0} \right] 
 \frac{n D_0 k^2 c(k) F(k;t)}{2}
 \nonumber \\ &&
+ \frac{D_0 k^2 S(0)}{3 S(k)^2} \partial_k F(k;t) \left[ \frac{d S(k)}{dk} - \frac{S(k)}{k}\right]
\nonumber \\ && 
+ \frac{D_0 k^2 S(0)}{6 S(k)^2} F(k;t) \left[  \frac{5}{k} \frac{d S(k)}{dk} - 
\frac{2}{S(k)} \left[ \frac{d S(k)}{dk} \right]^2
\right. \nonumber \\ && \left.
 + \frac{d^2 S(k)}{dk^2} \right]
 + \frac{D_0 k^2 S(0) n c(k)}{6} \partial^2_k F(k;t)
\nonumber \\ &&
+\frac{1}{3} \left[  \left. \frac{d^2 S(q)}{dq^2} \right|_{q=0} - \frac{S(0)}{k^2} \right] 
\int_0^t dt^\prime M^{\mathrm{irr}}(k;t-t^\prime) \partial_{t^\prime} F(k;t^\prime)
\nonumber \\ && 
+ \frac{1}{6} \frac{S(0)}{k}  \int_0^t d t^\prime M^{\mathrm{irr}}(k;t-t^\prime)  
\partial_{t^\prime} \partial_k F(k;t^\prime)
\nonumber \\ &&
+ \frac{S(0)}{6}  \int_0^t d t^\prime M^{\mathrm{irr}}(k;t-t^\prime)  
\partial_{t^\prime} \partial^2_k F(k;t^\prime)
\nonumber,
 \end{eqnarray}
and
\begin{eqnarray}
M_2^\chi(k;t) &=& n D_0  
\int \frac{d \mathbf{k}_1}{(2 \pi)^3} [v_{\mathbf{k}}(\mathbf{k}_1,\mathbf{k}-\mathbf{k}_1)]^2 
\nonumber \\ && \times F(|\mathbf{k}-\mathbf{k}_1|;t)  \chi^{(2)}(k_1;t),
\end{eqnarray}
and
\begin{eqnarray}\label{eq:M3}
\lefteqn{ M_3^\chi(k;t)  = } \\ \nonumber && n D_0 
\int \frac{d \mathbf{k}_1}{(2 \pi)^3} v_{\mathbf{k}}(\mathbf{k}_1,\mathbf{p}) F(p;t)  \chi^{(1)}(k_1;t)
\\ &&\nonumber \times
\left[ \frac{v_{\mathbf{k}_1}(\mathbf{k}_1,\mathbf{p})}{k} - \frac{2(\mathbf{k}\cdot\mathbf{k}_1)}{k^2 k_1}
 v_{\mathbf{k}}(\mathbf{k}_1,\mathbf{p}) + \frac{\mathbf{k}\cdot\mathbf{k}_1}{k} \frac{d c(k_1)}{d k_1} \right]
\\ \nonumber &&
 +\frac{n D_0}{2}  
\int \frac{d \mathbf{k}_1}{(2 \pi)^3} v_{\mathbf{k}}(\mathbf{k}_1,\mathbf{p}) F(p;t) 
 \chi^{(0)}(k_1;t)  
 \nonumber \\ && \times 
 \left[ \frac{c(k_1)}{k} - \frac{v_{\mathbf{k}}(\mathbf{k}_1,\mathbf{p})}{k^2}  
 + \frac{\mathbf{k} \cdot \mathbf{k}_1}{2 k} \frac{d^2c(k_1)}{d k_1^2}
 \right. \nonumber \\ && \nonumber \left.
 -\frac{2 (\mathbf{k}\cdot\mathbf{k}_1)^2 - 2 k^2 \mathbf{k}\cdot\mathbf{k}_1 - k^2 k_1^2}{k^3 k_1} 
\frac{d c(k_1)}{dk_1} 
 \right].
\end{eqnarray}
In Eq. \eqref{eq:M3} $\mathbf{p}=\mathbf{k}-\mathbf{k}_1$ and $p=|\mathbf{k}-\mathbf{k}_1|$. The 
initial condition is given by 
\begin{eqnarray}
\chi^{(2)}(k;0)  & = &  
 \frac{S^2(k)}{2}  \left[ \left. \frac{d^2 S(q)}{dq^2} \right|_{q=0} \right]
 \\ &&
+ \frac{1}{3} \frac{S(k) S(0)}{k} \frac{d S(k)}{dk} 
+ \frac{1}{6} S(k) S(0) \frac{d^2 S(k)}{dk^2} \nonumber
\end{eqnarray}

To solve Eq.~\ref{eq:chitwo} we also need the equation of motion \eqref{eq:dkF} for $\partial_k F(k;t)$ and 
the equation of motion for the second partial derivative of the intermediate scattering function with
respect to the wave-vector, $\partial^2_k F(k;t)$. The latter equation 
can be obtained from the mode-coupling theory equation of motion, Eq.~\eqref{eq:mct}:
\begin{eqnarray}
&&\partial_t \partial^2_k F(k;t) + \left[ \frac{ 2 D_0 }{S(k)} 
- \frac{4 D_0 k}{S(k)^2} \frac{d S(k)}{d k} \right] 
F(k;t)  
\nonumber \\ && 
-\left[ \frac{2}{S(k)} \left(\frac{d S(k)}{dk} \right)^2 + \frac{d^2S(k)}{dk^2} \right]
\frac{D_0 k^2}{S(k)^2}F(k;t) 
\nonumber \\ &&
+ \left[ \frac{4 D_0 k}{S(k)} - \frac{2 D_0 k^2}{S(k)^2} \frac{d S(k)}{dk} \right] \partial_k F(k;t)
\nonumber \\ &&
+ \frac{D_0 k^2}{S(k)} \partial^2_k F(k;t) 
\nonumber \\ &&
+ \int_0^t dt^\prime M^{\mathrm{irr}}(k;t-t^\prime) \partial_{t^\prime} \partial^2_k F(k;t^\prime)
\nonumber \\ &&
+\int_0^t dt^\prime M_1^{k2}(k;t-t^\prime) \partial_{t^\prime} \partial_k F(k;t^\prime) 
\nonumber\\&&
+\int_0^t dt^\prime M_2^{k2}(k;t-t^\prime) \partial_{t^\prime} F(k;t^\prime)
\nonumber \\ &=& 0,
\end{eqnarray}
where
\begin{eqnarray}\label{eq:M1k2}
\lefteqn{M_1^{k2}(k;t) =} \\ \nonumber && 
2 n D_0 \int \frac{d \mathbf{k}_1}{(2 \pi)^3} v_{\mathbf{k}}(\mathbf{k}_1,\mathbf{p})  
\left\{ c(p) + \frac{[\mathbf{k} \cdot \mathbf{p}]^2}{k^2 p}  
\frac{d c(p)}{d p} \right\} 
\nonumber \\ && \times 
F(k_1;t) F(p;t),
\nonumber \\ && \nonumber
+n D_0 \int \frac{d \mathbf{k}_1}{(2 \pi)^3} 
[v_{\mathbf{k}}(\mathbf{k}_1,\mathbf{p})]^2  
\frac{\mathbf{k} \cdot \mathbf{p}}{k p} 
\partial_{p} F(p;t-t^\prime) F(k_1;t)
\end{eqnarray}
and
\begin{eqnarray}\label{eq:M2k2}
\lefteqn{M_2^{k2}(k;t)  = } \\ \nonumber && 
n D_0 \int \frac{d \mathbf{k}_1}{(2 \pi)^3}
 F(k_1;t) F(p;t) 
\left[ c(p) + \frac{[\mathbf{k} \cdot \mathbf{p}]^2}{k^2 p}  
\frac{d c(p)}{d p} \right]^2
\nonumber \\ && \nonumber
+ n D_0 \int \frac{d \mathbf{k}_1}{(2 \pi)^3} v_{\mathbf{k}}(\mathbf{k}_1,\mathbf{p})  F(k_1;t) F(p;t)
\nonumber \\ && \times \nonumber
\left[ \frac{3 \mathbf{k} \cdot \mathbf{p}}{k p} \frac{d c(p)}{dp}
-\frac{[\mathbf{k}\cdot\mathbf{p}]^3}{k^3 p^3} \frac{dc(p)}{dp}
+\frac{[\mathbf{k}\cdot \mathbf{p}]^3}{k^3 p^2} \frac{d^2c(p)}{dp^2}
\right]
\\ \nonumber &&
+ 2 n D_0 \int \frac{d \mathbf{k}_1}{(2 \pi)^3}  v_{\mathbf{k}}(\mathbf{k}_1,\mathbf{p}) 
\left[ c(p) + \frac{[\mathbf{k} \cdot \mathbf{p}]^2}{k^2 p}  
\frac{d c(p)}{d p} \right]
\nonumber \\ && \nonumber
\times F(k_1;t) \frac{\mathbf{k}\cdot \mathbf{p}}{k p} 
\partial_p F(p;t)
\\ \nonumber &&
+\frac{n D_0}{2} \int \frac{d \mathbf{k}_1}{(2 \pi)^3} 
[v_{\mathbf{k}}(\mathbf{k}_1,\mathbf{p})]^2
\left[
\frac{1}{p} 
-\frac{[\mathbf{k} \cdot \mathbf{p}]^2}{k^2 p^3}
\right]
\nonumber \\ && \nonumber \times
\partial_p F(p;t) F(k_1;t)
\nonumber \\ && \nonumber 
+ \frac{n D_0}{2} \int \frac{d \mathbf{k}_1}{(2 \pi)^3} 
[v_{\mathbf{k}}(\mathbf{k}_1,\mathbf{p})]^2 
\left[ \frac{\mathbf{k} \cdot \mathbf{p}}{k p}\right]^2  
F(k_1;t)  \partial^2_p F(p;t).
\end{eqnarray}
In Eqs. (\ref{eq:M1k2}-\ref{eq:M2k2}) $\mathbf{p}=\mathbf{k}-\mathbf{k}_1$ and $p=|\mathbf{k}-\mathbf{k}_1|$.

\section{Isotropic Approximation}\label{sec:isotropic}

Along with the expansion of the full equation of motion \eqref{eq:motion}, 
we also examined an expansion of an isotropic approximation to Eq.~\eqref{eq:motion}. The isotropic approximation
has the advantage of being slightly easier computationally, and it allows for calculation of the
susceptibility at any $q$. 

The isotropic approximation assumes that the susceptibility  $\chi_{\mathbf{q}}(\mathbf{k};t)$ is 
independent of the angle between $\mathbf{q}$ and $\mathbf{k}$, 
$\chi_{\mathbf{q}}(\mathbf{k};t)\approx \chi_q^{\mathrm{iso}}(k;t)$. To derive an equation of motion 
for $\chi_q^{\mathrm{iso}}(k;t)$ one could start by substituting the isotropic approximation into the 
full equation of motion and then average the resulting equation over the angle between $\mathbf{q}$ 
and $\mathbf{k}$.
We propose a slight modification of this procedure that results in an equation that is somewhat easier
computationally:
\begin{eqnarray}\label{eq:isotropic}
&&\partial_t \chi_q^{\mathrm{iso}}(k;t) 
+\frac{D_0 k^2}{S(k)} \chi_q^{\mathrm{iso}}(k;t) 
\nonumber \\ && 
+\int_0^t dt^\prime M^{\mathrm{irr}}(k;t-t^\prime) \partial_{t^\prime} 
\chi_q^{\mathrm{iso}}(k;t^\prime)\nonumber \\
&&+ \int_0^t dt^\prime M^{\mathrm{iso}}_q(k;t-t^\prime)
\partial_{t^\prime} \tilde{F}(k;q;t^\prime)\nonumber\\
&=& 
n D_0 k^2 S(0) c(k) F(k;t) 
\nonumber \\ &&
+ S(0) \int_0^t dt^\prime M^{\mathrm{irr}}(k;t-t^\prime) \partial_{t^\prime} F(k;t^\prime).
\end{eqnarray}
where 
\begin{eqnarray}
M^{\mathrm{iso}}_q(k;t) & = & 
n D_0
\int \frac{d \mathbf{k}_1}{(2 \pi)^3}  \chi_q^{\mathrm{iso}} (k_1;t) F(|\mathbf{k}-\mathbf{k}_1|;t) 
\nonumber \\
&& \times v_{\mathbf{k}}(\mathbf{k}_1,\mathbf{k}-\mathbf{k}_1)
\tilde{v}_{\mathbf{k}}(\mathbf{k}_1,\mathbf{k}-\mathbf{k}_1;q)
\end{eqnarray}
and
\begin{eqnarray}
\tilde{F}(k;q;t) = \int \frac{d\hat{\mathbf{q}}}{4\pi} F(|\mathbf{k}+\mathbf{q}|;t),
\end{eqnarray}
\begin{eqnarray}
\tilde{v}_{\mathbf{k}}(\mathbf{k}_1,\mathbf{k}-\mathbf{k}_1;q) = \int \frac{d\hat{\mathbf{q}}}{4\pi} 
\frac{k v_{\mathbf{k}+\mathbf{q}}(\mathbf{k}_1+\mathbf{q},\mathbf{k}-\mathbf{k}_1)}{|\mathbf{k}+\mathbf{q}|}.
\end{eqnarray}
Finally, the initial condition for $\chi_q^{\mathrm{iso}}(k;t)$ is 
\begin{eqnarray}
\chi_q^{\mathrm{iso}}(k;t=0) = S(k) S(q) \int \frac{d\hat{\mathbf{q}}}{4\pi} S(|\mathbf{k}+\mathbf{q}|).
\end{eqnarray} 

Note that in Eq.~\eqref{eq:isotropic} we took the source term in the $q\to 0$ limit. 
The $q$ dependence of the source term has very little effect on the
size of the correlation length (see discussion in Sec.~\ref{sec:length}). Taking the source term
in the $q\to 0$ limit makes the numerical calculation of $\chi_q^{\mathrm{iso}}(k;t)$ 
somewhat easier.

To get the characteristic length we expanded $\chi_q^{\mathrm{iso}}(k;t)$ in powers of $q$,
\begin{eqnarray}
\chi^{\mathrm{iso}}_q(k;t) & = & \chi^{(0)}(k;t) + \frac{q^2}{2} 
\left[ \frac{\partial^2\chi_q^{\mathrm{iso}}(k;t)}{\partial q^2} \right]_{q=0} + \ldots \nonumber \\ 
& = & \chi^{(0)}(k;t) + q^2 \chi^{\mathrm{iso}(2)}(k;t)  + \ldots.
\end{eqnarray}

The zeroth order coefficient, $\chi^{(0)}(k;t)$, is the same as the one obtained from the expansion
of the complete equation of motion. The equation of motion for $\chi^{\mathrm{iso}(2)}(k;t)$ can be readily 
obtained from Eq. \eqref{eq:isotropic}:
\begin{eqnarray}\label{isoexpand1}
&&\partial_t \chi^{\mathrm{iso}(2)}(k;t) + \frac{D_0 k^2}{S(k)} \chi^{\mathrm{iso}(2)}(k;t) 
\nonumber \\ &&
+ \int_0^t dt^\prime M^{\mathrm{irr}}(k;t-t^\prime) \partial_{t^\prime} \chi^{\mathrm{iso}(2)}(k;t) \nonumber \\
&& + \int_0^t dt^\prime M^{\mathrm{iso}}_2(k;t-t^\prime) \partial_{t^\prime} F(k;t^\prime) \nonumber \\
&=& - \frac{1}{6} \int_0^t dt^\prime 
M^\chi_0(k;t-t^\prime)
\partial_{t^\prime} \partial^2_k F(k;t^\prime)   
 \nonumber \\ && 
 -\frac{1}{3k} \int_0^t dt^\prime 
M^\chi_0(k;t-t^\prime) \partial_{t^\prime} \partial_k F(k;t^\prime)  
 \nonumber \\ && 
- \frac{2}{3}\int_0^t dt^\prime M^{\mathrm{iso}}_3(k;t-t^\prime) \partial_{t^\prime} F(k;t^\prime)
\end{eqnarray}
where 
\begin{eqnarray}
\lefteqn{M^{\mathrm{iso}}_2(k;t)  = } \\ \nonumber && \nonumber 
n D_0 \int \frac{d \mathbf{k}_1}{(2 \pi)^3} [v_{\mathbf{k}}(\mathbf{k}_1,\mathbf{p})]^2
\chi^{\mathrm{iso}(2)}(k_1;t) F(|\mathbf{k} - \mathbf{k}_1|;t),
\end{eqnarray}
$M^\chi_0(k;t)$ is defined in Eq. \eqref{eq:Mchi0}, and
\begin{eqnarray}\label{eq:M3iso}
\lefteqn{M^{\mathrm{iso}}_3(k;t)  = } \\ && \nonumber 
\frac{n D_0}{2}  
\int \frac{d \mathbf{k}_1}{(2 \pi)^3} v_{\mathbf{k}}(\mathbf{k}_1,\mathbf{p}) F(p;t) 
 \chi^{(0)}(k_1;t)  
 \nonumber \\ && \times 
 \left[ \frac{c(k_1)}{k} - \frac{v_{\mathbf{k}}(\mathbf{k}_1,\mathbf{p})}{k^2}  
 + \frac{\mathbf{k} \cdot \mathbf{k}_1}{2 k} \frac{d^2c(k_1)}{d k_1^2}
 \right. \nonumber \\ && \nonumber \left.
 -\frac{2 (\mathbf{k}\cdot\mathbf{k}_1)^2 - 2 k^2 \mathbf{k}\cdot\mathbf{k}_1 - k^2 k_1^2}{k^3 k_1} 
\frac{d c(k_1)}{dk_1} 
 \right].
\end{eqnarray}
In Eq. \eqref{eq:M3iso} $\mathbf{p}=\mathbf{k}-\mathbf{k}_1$ and $p=|\mathbf{k}-\mathbf{k}_1|$.

The initial condition to Eq.~\eqref{isoexpand1} is given by
\begin{eqnarray}
\chi^{\mathrm{iso}(2)}(k;0) & = &
 \frac{S^2(k)}{2} \left[ \left. \frac{d^2 S(q)}{dq^2} \right|_{q=0} \right]
\\ && 
 + \frac{S(0) S(k)}{3k} \frac{dS(k)}{dk} 
+ \frac{S(0) S(k)}{6} \frac{d^2S(k)}{dk^2} \nonumber.
\end{eqnarray}

\section{Numerical evaluation of $\chi^{(n)}(k;t)$}\label{sec:numerical}
We numerically calculated the $k$ and $t$ dependence of $\chi^{(0)}(k;t)$, $\chi^{(1)}(k;t)$ and $\chi^{(2)}(k;t)$
using a previously developed algorithm that was designed to solve mode-coupling like 
equations \cite{Fuchs1991,Miyazaki,Flenner2005sim}. The only input in this calculation is the
static structure factor $S(k)$, which we calculated for the hard sphere interaction potential using the 
Percus-Yevick approximation. We report our results in terms of the relative distance from the ergodicity breaking 
transition predicted by mode-coupling theory, $\epsilon = (\phi_c - \phi)/\phi_c$. Here $\phi$ is the 
volume fraction, 
$\phi = n \pi \sigma^3/6$, where $\sigma$ is the hard sphere diameter, and $\phi_c$ is the volume
fraction at the mode-coupling transition. 
We used 300 equally spaced wave-vectors with spacing $\delta=0.2$, between $k_0=0.1$ and $k_{\mathrm{max}}=59.9$, 
and this discretization resulted in a mode-coupling transition at $\phi_c = 0.515866763$. 

\begin{figure}
\includegraphics[width=3.4in]{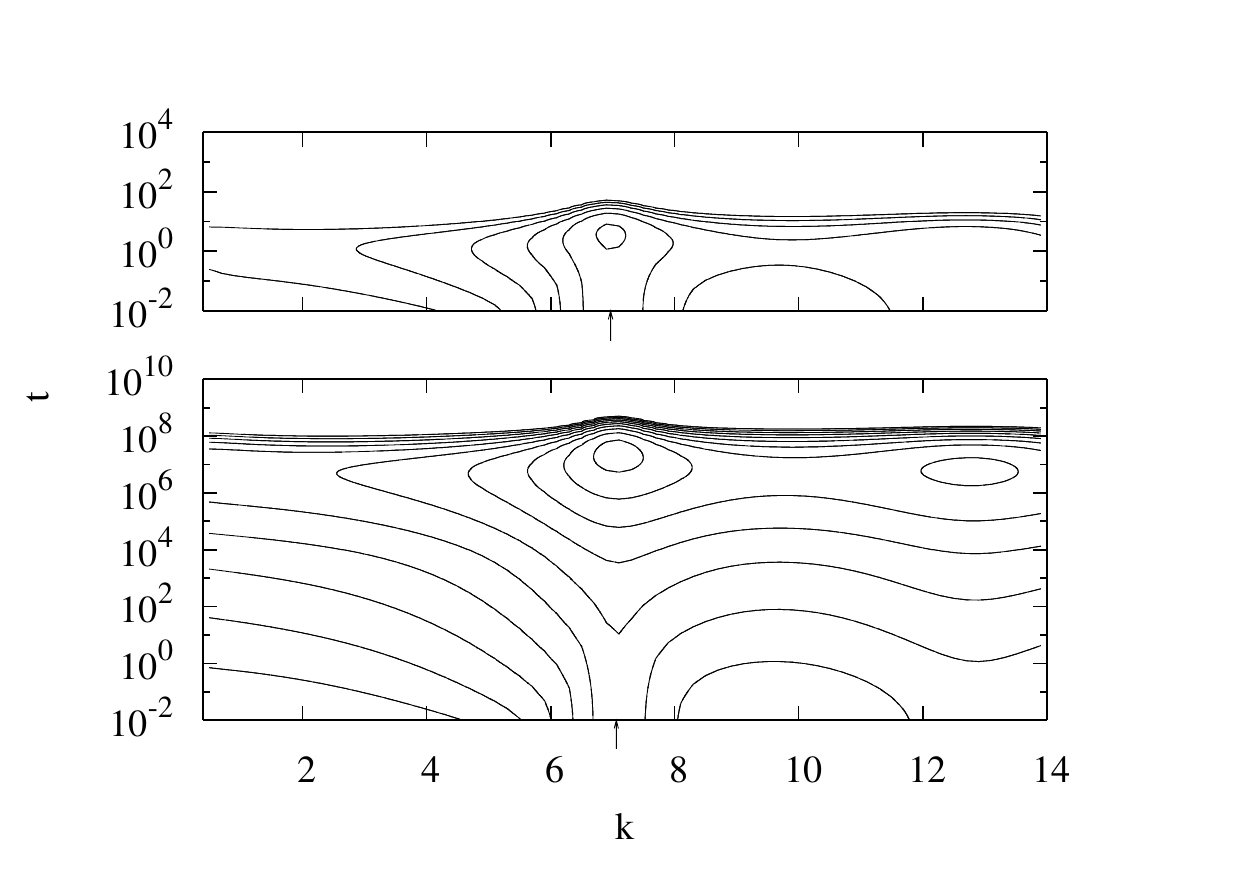}\\
\caption{\label{fig:contour}Time and wave-vector dependence of $\mathbf{q}=0$ value of 
three-point susceptibility, $\chi_{\mathbf{q}=0}(\mathbf{k};t)\equiv \chi^{(0)}(k;t)$ for
the reduced distance from the mode-coupling transition,
$\epsilon = (\phi_c-\phi)/\phi_c = 0.05$
(upper panel) and $\epsilon = 10^{-4}$ (lower panel).
Contours correspond to $\chi^{(0)}(k;t)= 4^n$ where $n$ is an integer, starting from $n=-6$.
The arrow marks the position of the first peak of the
static structure factor.}
\end{figure}

Shown in Fig.~\ref{fig:contour} are contour plots of $\chi^{(0)}(k;t)$ as a function of wave-vector $k$ and time 
$t$ for $\epsilon = 0.05$ and $\epsilon = 10^{-4}$. 
The former value of $\epsilon$ is the smallest relative distance 
from an avoided mode coupling transition in the Kob-Andersen binary mixture at which mode-coupling theory 
agrees with computer simulations \cite{Flenner2005sim}. 
As we mentioned earlier, $\chi^{(0)}(k;t)$ is proportional to the three-point susceptibility 
$\chi_n(k;t)$ calculated in Ref.\cite{Szamel2009}, which is a mode-coupling approximation for the density
derivative of the intermediate scattering function. Thus, all results derived in Ref.\cite{Szamel2009} for
for $\chi_n(k;t)$ also apply to $\chi^{(0)}(k;t)$. In particular, there is a well defined maximum in 
$\chi^{(0)}(k;t)$ at a well defined wave-vector and at a characteristic time. 
Also, all the scaling laws observed for $\chi_n(k;t)$ apply to $\chi^{(0)}(k;t)$ (we show some of these
scaling laws below). The characteristic wave-vector, $k_{\mathrm{max}}$, is nearly constant as the
mode coupling transition is approached and $k_{\mathrm{max}} \approx 7.1$ close to the transition.  

\begin{figure}
\includegraphics[width=3.1in]{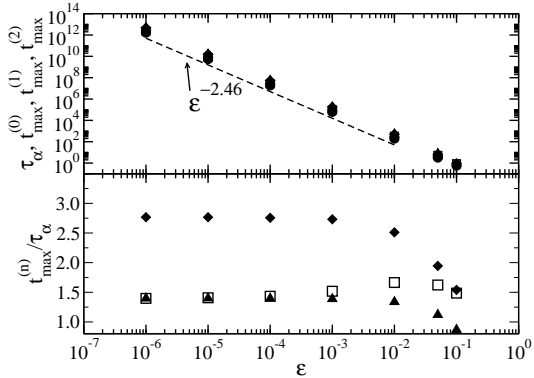}\\
\caption{\label{fig:time} Upper Panel: the $\alpha$ relaxation time, $\tau_\alpha$ (filled circles), and
the peak positions of $\chi^{(n)}(k_{\mathrm{max}};t)$, $\tau_{\mathrm{max}}^{(n)}$,
as a function of $\epsilon = (\phi-\phi_c)/\phi_c$: $\tau^{(0)}_{\mathrm{max}}$--triangles;
$\tau^{(1)}_{\mathrm{max}}$--diamonds; $\tau^{(2)}_{\mathrm{max}}$--open squares.
Lower Panel: the ratio $\tau_{\mathrm{max}}^{(n)}/\tau_\alpha$ as a function of $\epsilon$.}
\end{figure}

While the characteristic wave-vector is nearly constant close to the transition, the characteristic time grows 
rapidly as the mode-coupling transition is approached and diverges at the transition. 
In Fig.~\ref{fig:time} we examine the time at
which $\chi^{(n)}(k_{\mathrm{max}};t)$ is a maximum, $t_{\mathrm{max}}^{(n)}$, as a function
of $\epsilon$ for the characteristic wave-vector $k_{\mathrm{max}} = 7.1$. 
We compare  $t_{\mathrm{max}}^{(n)}$ with the $\alpha$ relaxation time $\tau_\alpha$, for which we use 
the standard definition $F(k_{\mathrm{max}};\tau_{\alpha}) = e^{-1}$. 
We find that $t_{\mathrm{max}}^{(n)}$ is slightly 
larger than $\tau_\alpha$, but has the same $\epsilon$ dependence, 
\textit{i.e.}\ $t_\mathrm{max}^{(n)} \sim \epsilon^{2.46}$.
Shown in the Fig.~\ref{fig:time}b are the ratios $t_{\mathrm{max}}^{(n)}/\tau_\alpha$, and 
it can be seen that these ratios are constant close to the mode-coupling transition. Thus,
in the $\epsilon\to 0$ limit we see that
$t_{\mathrm{max}}^{(0)} = 1.4 \tau_\alpha$, $t_{\mathrm{max}}^{(1)} =  2.8 \tau_\alpha$,
and $t_{\mathrm{max}}^{(2)} =1.4 \tau_\alpha$. Notice that the peak of
$\chi^{(0)}(k_\mathrm{max};t)$ and $\chi^{(2)}(k_\mathrm{max};t)$ occur at the same time.

Finally, in Fig.~\ref{fig:peak} we compare the values of $\chi^{(n)}(k_{\mathrm{max}};t)$ 
at $t_{\mathrm{max}}^{(n)}$.
We find that $|\chi^{(0)}(k_{\mathrm{max}};t_{\mathrm{max}}^{(0)})|$ and 
$|\chi^{(1)}(k_{\mathrm{max}};t_{\mathrm{max}}^{(1)})|$ grow as $\epsilon^{-1}$ whereas 
$|\chi^{(2)}(k_{\mathrm{max}};t_{\mathrm{max}}^{(2)})|$ grows as $\epsilon^{-3/2}$. As we discuss in the
next section, this disparate behavior of $\chi^{(n)}(k_{\mathrm{max}};t)$ is important for the existence 
of a diverging 
characteristic length.

We should note at this point that the $\epsilon$ dependence of $\chi^{(n)}(k;t)$ can be deduced from
scaling predictions described in Ref. \cite{Biroli2006}, and the numerical results presented here fully
agree with the these predictions.

\section{Diverging characteristic length}\label{sec:length}

To obtain a growing characteristic 
length scale as the mode-coupling transition is approached, we need $|\chi^{(n)}(k;t)|$  
for some $n > 0$ to grow faster than $|\chi^{(0)}(k;t)|$ at a fixed time $t$. 
Then a diverging length can be calculated as $|\chi^{(n)}(k;t)/\chi^{(0)}(k;t)|^{(1/n)}$. 

\begin{figure}
\includegraphics[width=3.1in]{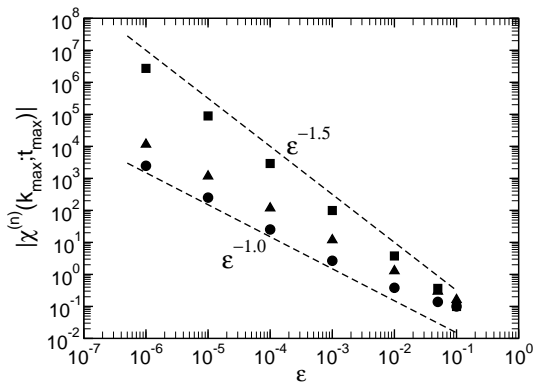}
\caption{\label{fig:peak}The peak height of $\chi^{(n)}(k;t)$ as
a function of the distance from the mode coupling transition $\epsilon$.
$\chi^{(0)}(k_{\mathrm{max}};t_{\mathrm{max}})$ -- triangles; 
$\chi^{(1)}(k_{\mathrm{max}};t_{\mathrm{max}})$ -- circles;
$\chi^{(2)}(k_{\mathrm{max}};t_{\mathrm{max}})$ -- squares}
\end{figure}

From Fig.~\ref{fig:peak} it is clear that the linear term, $\chi^{(1)}(k;t)$, does not result in 
a growing length scale. On the other hand, the absolute value of the isotropic second order term, 
$|\chi^{(2)}(k;t)|$, grows faster than $\chi^{(0)}(k;t)$ and thus we can define a diverging characteristic
length $\xi(k;t)$, 
\begin{equation}\label{eq:xi}
\xi(k;t) = \sqrt{-\frac{\chi^{(2)}(k;t)}{\chi^{(0)}(k;t)}},
\end{equation}
where the negative sign comes from the observation
that $\chi^{(2)}(k;t)$ is of opposite sign of $\chi^{(0)}(k;t)$ around $\tau_\alpha$ and close to the transition.
Note that for large times $t$, Eq.~\ref{eq:xi} involves a division of a small number by another small number. 
Because of
numerical issues present in the algorithm to calculate $\chi^{(n)}(k;t)$, 
we only show results if $\chi^{(n)}(k;t) \ge 10^{-3}$,
and therefore we, unfortunately, cannot comment at the asymptotic 
$t\to \infty$ limit of the characteristic length. 

\begin{figure}
\includegraphics[width=3.1in]{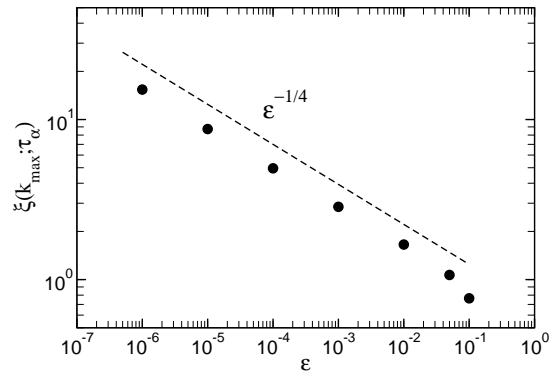}
\caption{\label{fig:length}The characteristic dynamic length $\xi(k_{\mathrm{max}};\tau_\alpha)$
as a function of the distance from the transition $\epsilon$.}
\end{figure}

In Fig.~\ref{fig:length} we examine $\xi(k_{\mathrm{max}};\tau_\alpha)$, \textit{i.e.} 
the characteristic length at $k=k_{\mathrm{max}}$ and at the $\alpha$ relaxation time. The
length $\xi(k_{\mathrm{max}};\tau_\alpha)$ grows as $\epsilon^{-1/4}$ and it reaches only 15 particle
diameters at $\epsilon = 10^{-6}$. Thus the characteristic length 
is not very large even very close to the transition. For
$\epsilon = 0.05$, we find that $\xi(k_{\mathrm{max}};\tau_\alpha)$ is only about one particle diameter.
Note that Eq.~\eqref{eq:xi} defines a length scale for every wave-vector $k$ and at all times $t$, 
and we examine the time and wave-vector dependence of $\xi(k;t)$ below.
 
We determined that setting the initial condition for $\chi^{(2)}(k;t=0)$ to zero and/or taking 
$\mathcal{S}^{(2)}(k;t) = 0$
had very little effect on the size of the correlation length close to the mode-coupling transition.
While including these terms is in principle straightforward, dropping them significantly 
simplifies the numerical calculation.

Fourier transforms of four-point correlation functions, \textit{i.e.} four-point dynamic structure
factors, are often monitored in simulations and used to investigate properties of dynamic heterogeneities.
Since the $q=0$ value of a four-point structure factor should be proportional to the characteristic
volume in which correlated motion takes place, an increase of the $q=0$ value (\textit{i.e.} of the 
height of four-point structure factor) is often given as evidence of an increase in a dynamic correlation length.

Similarly, for the problem considered here, the value of $\chi^{(0)}(k;t)$ could used as 
an indicator of the size of a characteristic dynamic range of the response. 
However, the spatial extent of dynamic response 
is best measured by examining the long-range spatial decay of a direct space susceptibility or,
alternatively, by examining the small-$q$ behavior of the susceptibility $\chi_{\mathbf{q}}(\mathbf{k};t)$.
This distinction is significant in view of the very strong wave-vector and time dependence 
of $\chi^{(0)}(k;t)$. In particular, if the characteristic length were a monotonic function of 
$\chi^{(0)}(k;t)$, then Fig.~\ref{fig:contour} would be leading to the unfortunate 
conclusion that $\xi(k;t)$ is a very strong function of $k$. The length would then have a rather limited appeal.
In the following paragraph we show that this is not the case.

\begin{figure}
\includegraphics[width=3.1in]{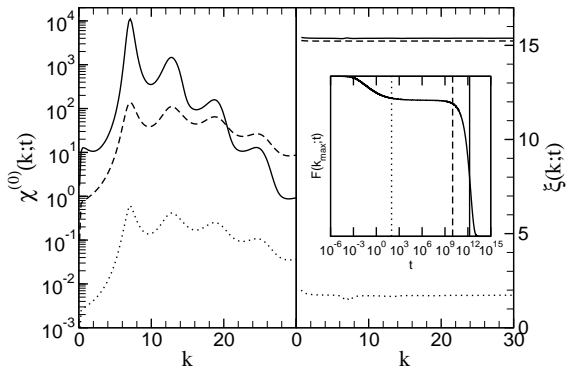}
\caption{\label{fig:chizlengthk} Left panel: the wave-vector dependence of the $\mathbf{q}=0$ value of
the three-point susceptibility, $\chi_{\mathbf{q}=0}(\mathbf{k};t)\equiv \chi^{(0)}(k;t)$, at a time
corresponding the the early $\beta$ relaxation regime (dotted line), the late $\beta$ regime 
(dashed line), and the $\alpha$ relaxation time (solid line). Right panel: the 
wave-vector dependence of the characteristic dynamic length $\xi(k;t)$ for the same times as in the left panel.
The inset is the self-intermediate scattering function 
$F(k;t)$ and the three vertical lines correspond to the three times
in left and right panels.}
\end{figure}

In Fig.~\ref{fig:chizlengthk}
we compare the $k$ dependence of  $\xi(k;t)$ (right figure) and $\chi^{(0)}(k;t)$ (left figure) 
for three characteristic times: (1) early $\beta$ (dotted line), late-$\beta$ (dashed line),
and at the $\alpha$ relaxation time (solid line). For reference, $F(k_{\mathrm{max}};t)$
is shown in the insert to Fig.~\ref{fig:chizlengthk} with the three characteristic times shown
as vertical lines in the figure. There is a very strong dependence of $\chi^{(0)}(k;t)$ on $k$, but
$\xi(k;\tau_\alpha)$ is nearly constant at each time. Therefore, even though there is a strong
$k$ dependence of the three-point susceptibility, there is a 
well defined characteristic dynamic length $\xi(k;t)$ that is independent of
$k$ and only depends on the time $t$. This suggests that we could drop the explicit $k$ dependence of $\xi(k;t)$
and introduce a simplified notation $\xi(t)$.

\begin{figure}
\includegraphics[width=3.1in]{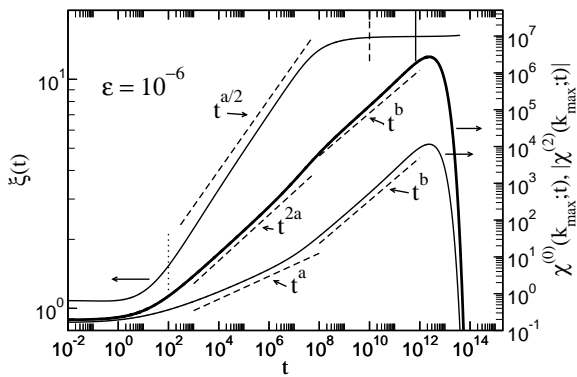}
\caption{\label{fig:chizlength}The time dependence of 
the characteristic dynamic length $\xi(t)$
(left solid line and left axis),
the susceptibility $\chi^{(0)}(k_{\mathrm{max}};t)$ (right solid line and right axis),
and the second order coefficient  $\chi^{(2)}(k_{\mathrm{max}};t)$ (middle, heavy solid line and right axis),
showed on a log-log scale.
The dashed lines show the scaling laws in the $\beta$ relaxation regime. The
vertical lines crossing $\xi(t)$ correspond to the three times shown in the inset to Fig.~\ref{fig:chizlengthk}.}
\end{figure}

Next, we investigate the time dependence of the characteristic length. 
Shown in Fig.~\ref{fig:chizlength} is $\chi^{(0)}(k_{\mathrm{max}};t)$ (lower curve-right axis), 
$|\chi^{(2)}(k_{\mathrm{max}};t)|$ (middle curve-right axis), and $\xi(t)$ 
(upper curve-left axis) 
as a function of time for $\epsilon = 10^{-6}$. The correlation length $\xi(t)$ is close to 
one for $t=0$, begins to grow during $\beta$ relaxation and reaches a plateau
at a time corresponding to the late $\beta$-early $\alpha$ relaxation. During the 
$\alpha$ relaxation, $\xi(t)$ is approximately constant. 
Note that $\xi(t)$ has a very different 
time dependence than $\chi^{(0)}(k_{\mathrm{max}};t)$. 
Therefore, the length scale associated with dynamic heterogeneities are 
not a maximum when $\chi^{(0)}(k;t)$ is a maximum, but rather reaches a constant value for
times less than this characteristic time. 

Scaling relations for different time regimes can be derived from the predictions
of the mode coupling theory \cite{Biroli2006,Berthier2007p2}. Specifically, in the early $\beta$ regime 
$\chi^{(0)}(k;t)\sim t^a$, and in the late $\beta$ regime
$\chi^{(0)}(k;t) \sim t^b$ where $a = 0.312$ and $b = 0.583$ for our system. 
The power law growth of $\chi^{(0)}(t)$ and $\chi^{(2)}(t)$ are also shown in Fig.~\ref{fig:chizlength}. During
the early $\beta$ relaxation regime, $\chi^{(0)}(t) \sim t^{a}$ while $\chi^{(2)}(t) \sim t^{2a}$,
which gives rise to the $t^{a/2}$ growth of the correlation length in the early $\beta$ relaxation regime.
However, during late $\beta$ relaxation, $\chi^{(2)}(t)$ and $\chi^{(0)}(t)$
both grow as $t^{b}$, thus there is no growing length scale.  
The vertical lines in the figure denote the same times as the vertical lines
in the inset to Fig.~\ref{fig:chizlengthk}.

\begin{figure}
\includegraphics[width=3.1in]{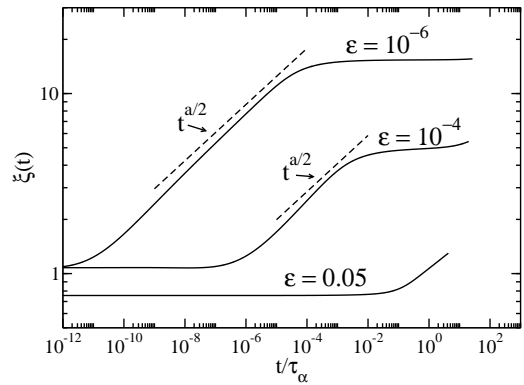}
\caption{\label{fig:timelength}The characteristic dynamic length $\xi(k_{\mathrm{max}};t)$
as a function of $t/\tau_\alpha$ for $\epsilon = 0.05$,
$10^{-4}$ and $10^{-6}$. The dashed lines is the scaling law $\xi(t) \sim t^{a/2}$ valid in the $\beta$ relaxation
regime.}
\end{figure}

In Fig.~\ref{fig:timelength} we show $\xi(k_{\mathrm{max}};t)$ 
as a function of $t/\tau_\alpha$ for $\epsilon=0.05$, $10^{-4}$, 
and $10^{-6}$. For $\epsilon = 10^{-4}$ we observe the $t^{a/2}$ scaling for only a very narrow range of time,
and we do not observe the $t^{a/2}$ scaling for any time range at $\epsilon = 0.05$, which suggests that
it might be very difficult to see this scaling in simulations. 

\begin{figure}
\includegraphics[width=3.1in]{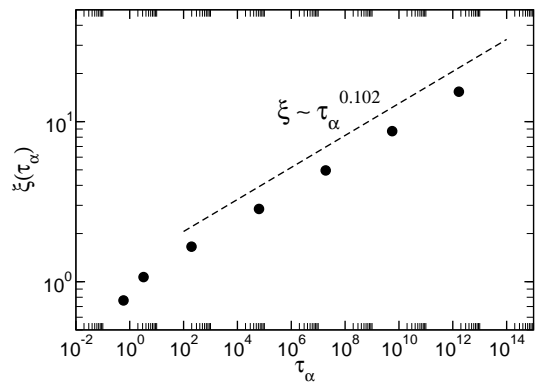}
\caption{\label{fig:lengthtau}The characteristic dynamic length $\xi(\tau_\alpha)$ calculated at the 
$\alpha$ relaxation time as a function of the $\alpha$ relaxation time.}
\end{figure}

Finally, we note that since
$\tau_\alpha \sim \epsilon^{-2.46}$ and $\xi \sim \epsilon ^{-0.25}$, then
$\xi \sim \tau_\alpha^{0.102}$, Fig.~\ref{fig:lengthtau}. As a result, a modest increase
in the correlation length is accompanied by a very large increase of the relaxation time.

\section{Numerical evaluation of $\chi_q^{\mathrm{iso}}(k;t)$ and associated characteristic length}
\label{sec:numericaliso}
The isotropic approximation neglects the dependence of the three-point
susceptibility on the angle between $\mathbf{k}$ and $\mathbf{q}$. Thus, in the resulting equation
of motion for $\chi_q^{\mathrm{iso}}(k;t)$ $q$ is just a parameter, and the equation of motion
can be solved separately for any value of $q$. As a result, 
the full $q$ dependence of $\chi_q^{\mathrm{iso}}(k;t)$ 
can be calculated. On the other hand, the isotropic approximation preserves the essential
terms in the equation of motion which lead to the divergence of the $q\to 0$ limit of 
$\chi_q^{\mathrm{iso}}(k;t)$ and of the characteristic length. 
In this section we examine the isotropic approximation
and compare this approximation to the expansion terms given above. Since the equations of
motion are similar and the terms that cause the divergence are identical, many of the
results of Sec.~\ref{sec:numerical} carry over to the isotropic approximation. Notably, as we already
noted in Sec. \ref{sec:isotropic}, $\chi^{(0)}(k;t)$ is identical in both cases. 

Since we can calculate the whole $q$ dependence in the isotropic approximation, we
can determine the characteristic length $\xi(t)$ using two different methods. We can either evaluate
$\chi_q^{\mathrm{iso}}(k;t)$ and then 
fit $\chi_{q}^{\mathrm{iso}}(k;t)/\chi^{(0)}(k;t)$ to $1- (\xi^{\mathrm{iso}}(k;t) q)^2$ 
for small $q$ or we can determine 
$\xi^{\mathrm{iso}}(k;t)$
from $\sqrt{-\chi^{iso(2)}(k;t)/\chi^{(0)}(k;t)}$. Both methods result in the same length.

It can be showed that within the isotropic approximation the characteristic length 
is almost $k$-independent (and thus we will denote it by $\xi^{\mathrm{iso}}(t)$). 
In addition, the time dependence of the length is very similar to what was
obtained from the full equations of motion in Sec.~\ref{sec:numerical}. 

\begin{figure}
\includegraphics[width=3.1in]{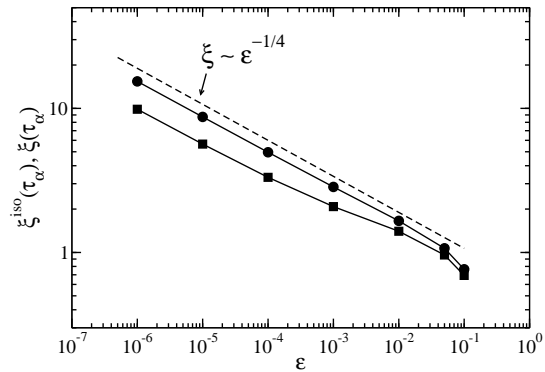}
\caption{\label{fig:lengthiso}The characteristic dynamic length $\xi^{\mathrm{iso}}(k;t)$ calculated using the
isotropic approximation (squares) and without the isotropic approximation (circles).}
\end{figure}

In Fig.~\ref{fig:lengthiso} we compare the magnitude of the characteristic length obtained from
the isotropic approximation, $\xi^{\mathrm{iso}}(\tau_\alpha)$, with that following from the full 
equations of motion, $\xi(\tau_\alpha)$. As we anticipated in the first paragraph of this section, 
the isotropic approximation gives a length which diverges as $\epsilon^{-1/4}$. However, the isotropic
approximation underestimates  the characteristic length; for small $\epsilon$ the length resulting
from the isotropic approximation is approximately 36\% smaller than the length resulting from the
expansion of the complete equation \eqref{eq:motion}.

There has been some discussion in the literature as to what scaling function should be used
to determine $\xi(t)$.  According to the scaling relation presented in Ref. \cite{Biroli2006}, 
in the $\beta$ and $\alpha$ regimes
the divergent part of $\chi_{\mathbf{q}}(\mathbf{k};t)$ is a function of a scaling variable
$q\xi(t)$ only for small $q$ close to the transition, 
$\chi_{\mathbf{q}}(\mathbf{k};t) = \mathcal{X}_{\beta,\alpha}(q\xi(t_{\beta,\alpha}),k)$.
We use the isotropic approximation to 
examine some properties of scaling function $\mathcal{X}_{\beta,\alpha}$ close to the 
mode coupling transition, in the $\beta$ and $\alpha$ regimes.

\begin{figure}
\includegraphics[width=3.1in]{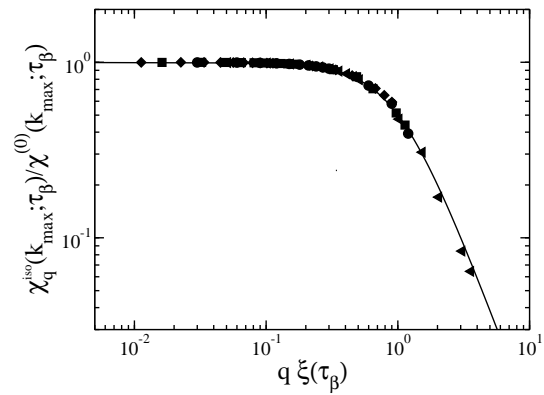}
\caption{\label{fig:qscalebeta}The isotropic approximation for the 
dynamic susceptibility $\chi_q^{\mathrm{iso}}(k_{\mathrm{max}};\tau_\beta)$ 
as a function of $q\xi(\tau_\beta)$ for $\epsilon \le 10^{-3}$. Only data for $q$ in the scaling 
regime are included.
The solid line is the Ornstein-Zernicke function $1/[1+(\xi q)^2]$.}
\end{figure}

For times $t$ in the vicinity of the $\beta$ relaxation time $\tau_\beta$, 
the scaling function $\mathcal{X}_\beta(q\xi(t_\beta),k)$ is predicted
to have the Ornstein-Zernicke behavior, namely $\mathcal{X}_\beta(q\xi(t_\beta),k)$ should 
scale as $q^{-2}$ for large $q$ \cite{Biroli2006}. To check this prediction we 
first need to define the $\beta$ relaxation time. We define $\tau_\beta$ as the
inflection point of $F(t)$ versus $\ln(t)$. We verified that this definition 
agrees with the MCT scaling $\tau_\beta \sim \epsilon^{-1/2a}$. This time $\tau_\beta$ is only well
defined for $\epsilon \le 10^{-3}$. 
Shown in Fig.~\ref{fig:qscalebeta} is 
$\chi_q^{\mathrm{iso}}(k_{\mathrm{max}};\tau_\beta)/\chi^{(0)}(k_{\mathrm{max}};\tau_\beta)$
as a function of $q \xi(\tau_\alpha)$ 
and the Ornstein-Zernicke function $1/[1+(\xi q)^2]$, which provides a good fit for 
small $q$ during the $\beta$ relaxation time and demonstrates the $q^{-2}$ scaling
for large $q$.

\begin{figure}
\includegraphics[width=3.1in]{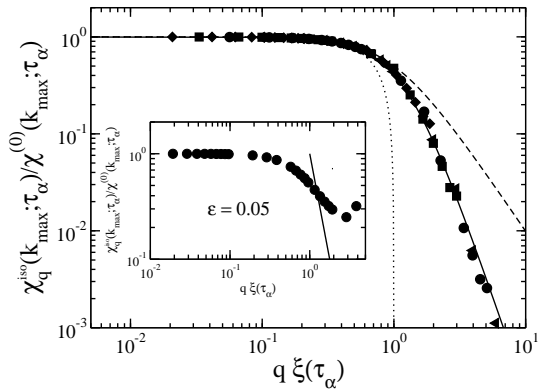}
\caption{\label{fig:qscale}The isotropic approximation for the 
dynamic susceptibility $\chi_q^{\mathrm{iso}}(k_{\mathrm{max}};\tau_\alpha)$ 
as a function of $q\xi(\tau_\alpha)$ for $\epsilon \le 10^{-3}$. Only data for $q$ in the scaling 
regime are included. The dotted line is $1-(q\xi)^2$, the dashed line is $1/[1+(q\xi)^2]$.
The solid line is a fit to the data to a function of the form $1/[1 + (q\xi)^2 + a (q\xi)^4]$
where $a=0.45$. The $q^{-4}$ scaling for large $q$ is evident. Inset: the dynamic 
susceptibility $\chi_q^{\mathrm{iso}}(k_{\mathrm{max}};\tau_\alpha)$ for $\epsilon = 0.05$ showing all the data
including $q$ values beyond the scaling regime. The inset shows that
the $q^{-4}$ scaling (solid line) is not apparent for this $\epsilon$.}
\end{figure}

For times $t$ comparable to the $\alpha$ relaxation time $\tau_\alpha$, the inhomogeneous mode-coupling
theory \cite{Biroli2006} predicts a $q^{-4}$ behavior of the scaling function 
$\mathcal{X}_\alpha(q\xi(t_\alpha),k)$at large $q$.
We test this prediction in Fig.~\ref{fig:qscale}: we show  
$\chi_{q}^{\mathrm{iso}}(k_{\mathrm{max}};\tau_\alpha)/\chi^{(0)}(k_{\mathrm{max}};\tau_\alpha)$ 
as a function of $q \xi(\tau_\alpha)$ along with two functions commonly used
to find $\xi(t)$ in simulations, and a function suggested by the inhomogeneous mode-coupling theory. 
The functions $1-(\xi q)^2$ (dotted line) and the Ornstein-Zernicke function, $1/(1+[\xi q]^2)$, (dashed line) 
are good fits only to a very narrow $q$ range, with the Ornstein-Zernicke 
function being a better fit for a larger range of $q$ values. On the other hand, 
the function $1/[1+(\xi q)^2 + a(\xi q)^4]$ where 
$a = 0.45$ (solid line), provides a good fit over a large $q$ range and thus it  
confirms the $q^{-4}$ scaling predicted by the inhomogeneous mode-coupling
theory for the $\alpha$ relaxation time scale. 
Note that the $q^{-4}$ scaling is not evident for $\epsilon = 0.05$ (inset),
which suggests that this scaling might be difficult to observe in simulations.  

\section{Conclusions}\label{sec:conclusions}
We used inhomogeneous mode-coupling theory to numerically investigate 
the dynamic susceptibility $\chi_{\mathbf{q}}(\mathbf{k};t)$ at small
$q$ and determined time, $k$, and distance from the transition dependence of the diverging
characteristic length scale. We confirmed scaling predictions presented 
in Ref. \cite{Biroli2006} and obtained a couple of new interesting results.
Because of numerical issues, we
were not able to calculate the asymptotic long time behavior of the 
diverging characteristic length scale. This would be an
interesting topic that we leave for later analysis. It most likely requires an analytical argument
that goes beyond the scaling analysis presented in Ref. \cite{Biroli2006}. 

The most important result of our numerical investigation is that the diverging characteristic length
is very weakly $k$ dependent. This makes it a well defined quantity. We speculate that the 
$k$ independence of the characteristic length should carry over to the dynamic correlation length
defined in terms of a four-point structure factor. Moreover, it should explain why a variety
of slightly different four point functions (\textit{e.g.} defined in terms 
of overlap functions \cite{Lacevic2003,Stein2008,Karmakar2009} or in terms of 
self-intermediate scattering functions \cite{Berthier2004,Toninelli2005,Flenner2009})
result in comparable dynamic correlation lengths. 

The second important result, which cannot be obtained from scaling considerations alone, is the magnitude
of the characteristic length. On general grounds we expect this length to be comparable to dynamic
correlation lengths defined through four-point structure factors. Thus, it is satisfying that the 
magnitude of the length is indeed comparable (albeit somewhat smaller) to what's found in simulations.

We would like to point out that, although various simulations found comparable values of the
dynamic correlation length, there are a few important unresolved differences between results obtained
by different groups that preclude declaring that
the characteristic length discussed in this work is essentially the same as the dynamic
correlation length measured in simulations.

First, while the characteristic length defined through the three-point susceptibility is 
a monotonic function of time (at least as long as our numerical routines are reliable), 
the simulational results very. Lacevic \textit{et al.} \cite{Lacevic2003} found that 
the dynamic correlation length roughly followed the overall 
magnitude of the four-point correlation function and decayed to zero at later times. In contrast,
Toninelli \textit{et al.} \cite{Toninelli2005} found
that the dynamic correlation length continued to grow at later times. While slightly different fitting
procedures were used in these two works, it is difficult to pinpoint the exact source of two
strikingly different results.

Second, within the mode-coupling approximation, the characteristic length defined through the three-point susceptibility 
diverges as $\epsilon^{-1/4}$ upon approaching the ergodicity breaking transition predicted by the 
mode-coupling theory. We feel  that the relevance of this result to simulations (and experiments) 
in which the mode-coupling transtion is avoided still 
needs to be  fully established. We speculate that it is possible that in computer simulations 
a vestige of a power law divergence of the dynamic correlation length could be seen just as one 
can observe in simulations power law dependencies of various transport coefficients 
upon approaching a mode-coupling crossover \cite{Flenner2005sim}.
Indeed, various groups have already claimed power law dependencies of their dynamic correlation
lengths upon approaching the mode-coupling crossover 
(see, \textit{e.g.} \cite{Berthier2004,Whitelam2004,Flenner2009,Stein2008,Karmakar2009}). 
However, there seems to be some disagreement regarding the value of the scaling exponent and only one work,
\cite{Stein2008}, results in a value agreeing with the prediction of the inhomogeneous
mode-coupling theory. Upon closer examination of the fitting procedure described in Ref. 
\cite{SteinPhD} and re-examining our own simulation data we concluded that virtually all
systems studied in simulations were not large enough to obtain the dynamic correlation length
in a range allowing for an unambiguous determination of the scaling exponent.

\section{Acknowledgment}
We would like to thank G. Biroli, K. Miyazaki and D. Reichman for comments about this work. 
We gratefully acknowledge the support of NSF Grant No.\ CHE 0517709.

\end{document}